# LITERATURE REVIEW OF THE EFFECT OF QUANTUM COMPUTING ON CRYPTOCURRENCIES USING BLOCKCHAIN TECHNOLOGY

**Adi Mutha**
**12th Standard Student, Dr. Kalmadi Shamarao Junior College**
adi.mutha7508@gmail.com

**Dr. Jitendra Sandu**
**Ceo**
**Talent Assessment & Analytics Software Solutions (TaaS)** Jitendra@FindYourFit.in

## 1. ABSTRACT

With the advent of quantum computing, cryptocurrencies that rely on blockchain technology face mounting cryptographic vulnerabilities. This paper presents a comprehensive literature review evaluating how quantum algorithms—specifically Shor's and Grover's—could disrupt the foundational security mechanisms of cryptocurrencies. Shor's algorithm poses a threat to public-key cryptographic schemes by enabling efficient factorization and discrete logarithm solving, thereby endangering digital signature systems. Grover's algorithm undermines hash-based functions, increasing the feasibility of 51% attacks and hash collisions.

By examining the internal mechanisms of major cryptocurrencies such as Bitcoin, Ethereum, Litecoin, Monero, and Zcash, this review identifies specific vulnerabilities in transaction and consensus processes. It further analyses the current hardware limitations of quantum systems and estimates when such attacks could become feasible. In anticipation, it investigates countermeasures including Post-Quantum Cryptography (PQC), Quantum Key Distribution (QKD), and protocol-level modifications such as memoryintensive proof-of-work algorithms and multi-signature schemes. The discussion integrates recent advancements in quantum error correction, hardware scalability, and NIST-standardized cryptographic algorithms.

This review concludes that while quantum computers are not yet advanced enough to pose an immediate threat, proactive integration of quantum-resistant solutions is essential. The findings underscore the urgent need for cryptocurrencies to adopt post-quantum cryptographic standards to preserve the decentralized trust, integrity, and security that define blockchain-based digital cryptocurrencies.

## 2. INTRODUCTION

Blockchain is a decentralised, digital ledger that records transactions across a network. It enables multiple remote and potentially untrusting participants to engage in transactions by relying on a protocol whose immutability and consensus mechanisms establish a foundational level of trust among them (Stephen Holmes & Liqun Chen, 2021).

Blockchain was first introduced in 2008 when an anonymous individual or group known as Satoshi Nakamoto published the Bitcoin white paper (Nakamoto, 2008). Since then, hundreds of other decentralised digital cryptocurrencies have been developed in a market currently worth over 3.46 trillion USD. These cryptocurrency systems are secured through the use of public key asymmetric cryptography, like RSA or Elliptic Curve (EC) encryption, and cryptographic hash functions like SHA-256, Scrypt, and RandomX.

Cryptocurrency systems face a growing threat from the advancement of quantum computing. Unlike classical computers, which operate on binary bits that exist in a single state—either 0 or 1—quantum computers utilize qubits, which can exist in multiple states simultaneously due to the principles of superposition and entanglement fundamental to quantum mechanics (Kappert, Karger, & Kureljusic, 2021). This allows quantum computers to perform parallel computations that have no counterpart in the classical world. Therefore, this enables the quantum computers to execute two of the most significant quantum algorithms: Shor's algorithm and Grover's algorithm.

Shor's algorithm, introduced in 1994 by Peter Shor, can factor large integers and solve discrete logarithmic problems in polynomial time (Shor, 1994). This algorithm performed by quantum computers provides it exponential speedup compared to classical ones. Consequently, quantum computers executing Shor's algorithm pose a significant threat to existing public-key cryptographic systems used by cryptocurrencies, as they enable the extraction of private keys from corresponding public keys. This capability undermines the integrity of digital signatures and facilitates a form of attack known as transaction hijacking, where unauthorized entities can forge transactions (Kappert, Karger, & Kureljusic, 2021).

On the other hand, Grover (1996) introduced another quantum algorithm that provides a quadratic speedup for solving unstructured search problems compared to classical approaches. This poses a threat to the hash functions used in blockchain systems, as it can be used to find the pre-image of a hash output efficiently. Such algorithmic power benefits the attacker in two significant ways (Kappert, Karger, & Kureljusic, 2021):

1. It enables the attacker to more easily search for hash collisions, potentially allowing them to replace blocks without compromising the chain's structural integrity.
2. It increases the feasibility of conducting a 51% attack during mining of the cryptocurrency, which is discussed in detail.

This literature review aims to comprehensively examine the potential threats that quantum algorithms pose to existing blockchain and cryptocurrency systems. It not only evaluates the feasibility of these quantum attacks but also assesses their practical implications, including computational cost, estimated time to execution, and the projected timeline for when quantum systems may realistically become capable of compromising cryptographic security. Furthermore, the review explores and compares various proposed post-quantum mitigation strategies, including hard forks, soft forks, and quantum-resistant cryptographic schemes. In this paper, we examine not only the theoretical implications but also the practical applications within the context of quantum threats to blockchain systems.

## 3. METHODOLOGY

This literature review adopts a structured approach to explore the impact of quantum computing on blockchain-based cryptocurrencies. The goal was to identify, analyse, and synthesize existing academic insights into the vulnerabilities posed by quantum algorithms and the proposed countermeasures.

A systematic search was conducted using academic databases including Google Scholar, IEEE Xplore, SpringerLink, and ScienceDirect. The primary search phrase used was “quantum computing in blockchain”. Supplementary keywords such as “Shor’s algorithm,” “Grover’s algorithm,” “post-quantum cryptography,” and “cryptocurrency security” were also used to refine the results.

Initially, 103 research papers were identified. These were shortlisted through a multi-stage filtering process:

Titles and abstracts were reviewed for relevance to the intersection of quantum computing and blockchain. Papers with higher citation counts and recent publication dates were prioritized.

Studies were included only if they provided technical depth or novel insights into cryptographic vulnerabilities, consensus mechanisms, or post-quantum solutions.

The final selection comprised 46 research papers. These were categorized using a self-developed Excel matrix, which divided the content into subtopics such as:

- Nature of quantum threats
- Targeted vulnerabilities in cryptocurrencies
- Quantum algorithms
- Countermeasures (PQC, QKD, protocol updates, etc.)
- Future prospects and limitations

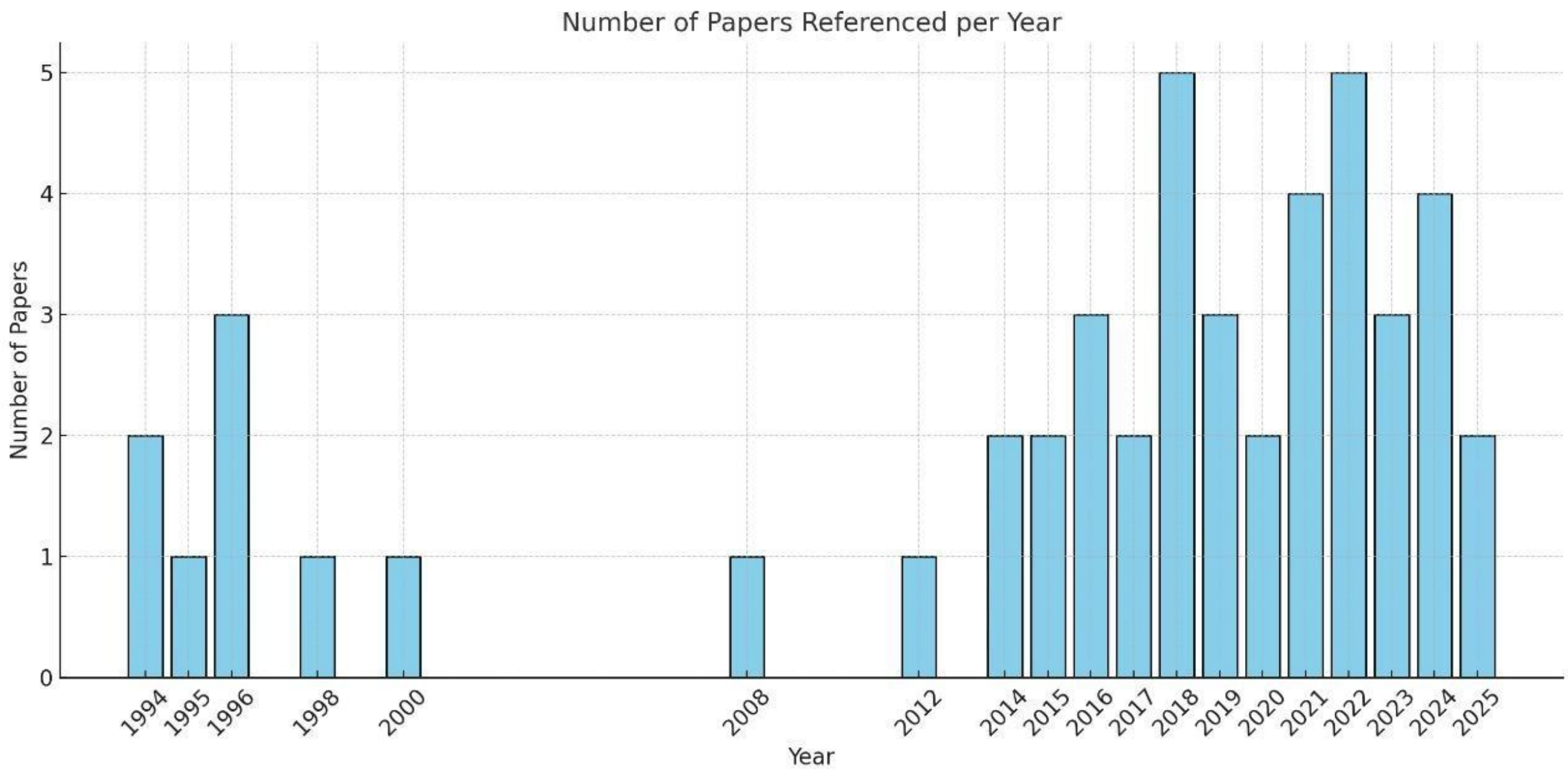

*Figure 1. Number of papers per year*

This structured methodology enabled a thematic synthesis of existing knowledge which highlights the complexity of transitioning to quantum-resilient cryptocurrency systems.

## 4. LITERATURE REVIEW

A literature review involves a thorough and analytical examination of previously published academic works, including research papers, books, and articles, that relate to a particular topic, research question, or field of inquiry.

Quantum computing, a cutting-edge interdisciplinary field emerging from the convergence of physics and computer science, has seen continuous growth since its conceptual foundation in the 1980s by physicist Richard Feynman. It poses a potential threat to the encryption of cryptocurrency systems. Therefore, this article will go into an in-depth analysis of the major security threats posed by quantum algorithms to various cryptocurrencies, their countermeasures, limitations, and future trends.

3.1.1 EXPLANATION AND WORKING

Before analysing the significant threats posed by quantum computing, it is necessary to understand basic concepts and working of blockchain, public key cryptography and quantum computing.

A. Blockchain and Cryptocurrencies

Blockchain is a distributed, decentralised public ledger that serves as the foundation for most major cryptocurrencies. Its primary application lies in enabling secure and transparent digital transactions while effectively addressing the issue of double-spending without the need for a central authority.

As outlined by Nakamoto (2008) in the Bitcoin whitepaper, unprocessed transactions are bundled together, which, after verifying and validating, form a block. Each block is assigned a timestamp, ensuring that all transactions are recorded in chronological order (Katz & Lindell, 2014). This chronological structure supports the integrity of the blockchain by preventing double-spending and enhancing resistance to tampering. Furthermore, each block is cryptographically linked to its predecessor through the hash of the previous block's header, forming a secure and immutable chain.

In the context of blockchain, a node refers to any computer or device that participates in the Bitcoin network. There are primarily three types of nodes: user nodes, which initiate and receive transactions; validator nodes, which verify and validate transactions; and mining nodes, which perform the consensus mechanism to validate unconfirmed transactions and append them to the blockchain (Fernández-Caramés & Fraga-Lamas, 2020).

Thus, the blockchain depends on 2 main computational mechanisms (Kearney & Perez-Delgado, 2021):-

1. Transaction mechanism
2. Consensus mechanism

The transaction mechanism in blockchain networks facilitates the transfer of cryptocurrencies and associated data between user nodes. This mechanism relies on public-key cryptography, specifically digital signature algorithm schemes such as ECDSA, RSA, and Schnorr signatures (Kearney & Perez-Delgado, 2021). Among these, the Elliptic Curve Digital Signature Algorithm (ECDSA) is predominantly used across most cryptocurrency systems due to its relatively small key size and robust security properties. These cryptographic schemes are asymmetric, meaning that the private key cannot be feasibly derived from the public key. During a transaction, the signature algorithm generates a unique public-private key pair of fixed length. In many blockchain systems, the public key is further hashed (typically using SHA-256 and RIPEMD-160) to produce a wallet address. Each transaction in the blockchain includes a reference to the previous block's hash, a timestamp, and other transaction-specific metadata, which vary depending on the specific cryptocurrency protocol. To initiate a transaction, the sender enters the recipient's wallet address and the amount to be transferred, then signs the transaction with their private key, generating a digital signature and verifying ownership of the wallet funds. This signed transaction is then propagated to the mempool, where it awaits validation. Mining nodes then verify the digital signature using the sender's public key, thereby confirming the sender's ownership of the associated funds. Once verified, the transaction is included in a block through the execution of a consensus mechanism (e.g., Proof of Work), which finalizes and records it on the blockchain.

The consensus mechanism serves a critical role in maintaining the integrity and consistency of blockchain networks. It ensures the prevention of double-spending, the validation and security of transactions, and the propagation and final inclusion of verified blocks into the blockchain ledger. These operations are primarily executed by mining nodes, which are incentivized through rewards or transaction fees—according to the protocols of their respective cryptocurrencies.
Mining nodes verify the legitimacy of transactions by examining whether the sender possesses sufficient funds. Since the blockchain records details of sender and receiver wallets along with transaction amounts, it enables verification of the Bitcoin balance associated with a specific wallet. This process resembles checking the balance of a conventional bank account, with the wallet address or public key serving as the equivalent of an account number (Segendorf, 2014).
On the other hand, the central function of these nodes involves solving complex cryptographic puzzles as part of the consensus algorithm to validate the transaction. For instance, in Bitcoin's Proof-of-Work mechanism, miners search for a nonce that results in a hash meeting a predefined difficulty threshold (Aggarwal et al., 2018). As the number of participating miners increases, the network dynamically adjusts the difficulty level to maintain consistent block generation times. Once a miner successfully solves the cryptographic puzzle, it is broadcast across the network for verification by other nodes. These nodes confirm whether the new block is a valid successor to the existing chain by evaluation of the hash function. Upon consensus, the block is appended to the blockchain, ensuring transparency and immutability of the ledger (Vallois & Guenane, 2017). The mining process requires substantial computational resources and relies on highly advanced ASIC-based systems to execute the vast number of calculations required. In return for this intensive computational effort, miners are rewarded for their work (Kearney & Perez-Delgado, 2021).

Major cryptocurrencies, such as Bitcoin, Ethereum, Litecoin, Monero, and Zcash, each implement distinct transaction models and consensus mechanisms, as explained in Table 2.

| **Cryptocurrency** | **Signature Algorithm** | **Public Key Size** | **Consensus Mechanism** | **Hash Function** | **Hash Signature Size** | Block Interval Time |
|---|---|---|---|---|---|---|
| **Bitcoin** | ECDSA (secp256k1) | 256 bits | Hashcash - PoW | SHA-256 | 256 bits and 160 bits | 10 mins |
| **Ethereum** | ECDSA (secp256k1) | 256 bits | Ethereum 2.0 - PoS | Keccak-256 | 256 bits | 15 seconds |
| **Litecoin** | ECDSA (secp256k1) | 256 bits | Scrypt - PoW | Scrypt | 256 bits | 2.5 mins |
| **Monero** | EdDSA (Ed25519) + Ring Signatures | 256 bits | RandomX - PoW | Keccak-256 | 256 bits | 2 mins |
| **Zcash** | ECDSA (secp256k1) + zk-SNARKs | 256 bits | Equihash PoW | SHA-256 | 256 bits | 75 seconds |

Table 1 presents an overview of the selected cryptocurrencies, highlighting the signature algorithms employed as public key cryptography, their respective consensus mechanisms, hash function, and hash signature size (Fernández-Caramés & Fraga-Lamas, 2020).

*Table 1. Public key cryptosystems and hash functions used by various cryptocurrencies*

A hash function is a mathematical algorithm that transforms input data of any size into a fixed string of characters (usually 64-bit hexadecimal string). This output is unique to the input, which includes the Merkle root, the hash of the previous block, the timestamp, the nonce, etc. Even a minor alteration in the input results in a significantly different hash. As the block contains the hash of the previous block, any modification to the previous block will make the next block's hash invalid, and so the blockchain (Rodenburg and Pappas, 2017). According to Nakamoto's consensus protocol, the blockchain network recognizes the longest valid chain as the legitimate version of the distributed ledger (Nakamoto, 2008). Thus, a malicious attacker will have to recalculate all the hashes and perform the fastest PoW to maintain the longest chain against the entire mining network (Kappert, Karger, & Kureljusic, 2021). Consequently, for a malicious attacker to successfully introduce a fraudulent block, they would need to expend an immense amount of computational power to outperform honest nodes in the consensus process, not just for a single block, but continuously, to maintain the longest chain. While such an attack is highly improbable with classical computing due to the vast energy and resources required, the emergence of quantum computing poses a potential threat by making this level of computational dominance more feasible (Kearney & Perez-Delgado, 2021).

| Cryptocurrency | Transaction Mechanism | Consensus Mechanism |
|---|---|---|
| **Bitcoin (BTC)** | The sender signs the transaction with their private key and broadcasts it to the Bitcoin network. Nodes verify its authenticity and balance, after which it enters the mempool. A miner then includes it in a block and adds it to the blockchain. | **Proof of Work (PoW – Hashcash):** Miners solve a cryptographic puzzle (SHA-256 hash) to find a nonce that results in a block hash below a target. First, to solve broadcasts,the broadcast block. The network verifies and accepts the longest valid chain. Mining secures the network and issues new coins. |
| **Litecoin (LTC)** | The sender signs the transaction using their private key and broadcasts it to the peer-to-peer Litecoin network. Nodes verify the transaction against the UTXO set. Once validated, it enters the mempool and is later included in a new block by a miner. | **Proof of Work (PoW – Scrypt):** Similar to Bitcoin but uses the **Scrypt** algorithm instead of SHA-256. Scrypt is more memoryintensive, aiming for fairer mining and ASIC resistance. Block time is ~2.5 minutes. |
| **Ethereum (ETH)** | In Ethereum, a user signs a transaction including a gas fee and broadcasts it to the network. Nodes verify its validity, after which it enters the transaction pool. A validator then selects and includes it in a new block through the Proof of Stake consensus mechanism. | **Proof of Stake (PoS – Ethereum 2.0/Beacon Chain):** Validators are selected pseudorandomly to propose and attest to blocks. Staking 32 ETH is required to become a validator. Honest validators are rewarded, while malicious ones can be slashed. Finality is achieved via the Casper FFG protocol. |
| **Monero (XMR)** | The sender signs the transaction using a Ring Signature and broadcasts it to the network. Nodes verify the transaction without revealing the sender's identity. A stealth address conceals the recipient, and the amount is hidden using RingCT. The transaction is then mined into a block. | **Proof of Work (PoW – RandomX):** Optimized for CPUs to resist ASICs. RandomX dynamically generates a program and dataset per block to be executed by the miner, making specialized hardware less effective. Protects decentralization. |
| **Zcash (ZEC)** | The user creates a transaction, choosing between shielded or transparent options. For shielded transactions, a zk-SNARK proof is generated. The transaction is broadcast to the network, verified | **Proof of Work (PoW – Equihash):**<br>Memoryhard algorithm to prevent ASIC dominance.<br>Uses zero-knowledge proof validation.<br>Emphasis on privacy with transparent and |
| | through zero-knowledge proofs without revealing sensitive data, and finally mined into a block. | shielded pools. Miners still compete to solve Equihash puzzles. |

*Table 2. Transaction and consensus mechanism of each cryptocurrency*

B. Quantum Computing

Quantum computing, first proposed by Richard Feynman in 1982, represents a convergence of quantum physics and computer science. Unlike classical computers, which operate using binary bits restricted to a state of either 0 or 1, quantum computers utilize quantum bits, or 'qubits', which are the smallest units of information in quantum computing. The qubits are typically subatomic particles like electrons or photons which exist in multiple states simultaneously, which upon inspection collapse into one of the two states: 0 and 1 (Allende et al., 2023; Mavroeidis et al., 2018). This fundamental distinction enables quantum computers to perform certain computations more efficiently than their classical counterparts (Dey et al., 2022). Quantum computers can be broadly classified into two categories: universal and non-universal. In the context of blockchain and

Cryptocurrencies, non-universal quantum computers are particularly relevant, as they are specialized machines designed to perform specific tasks, such as breaking RSA encryption schemes (Mavroeidis et al.,

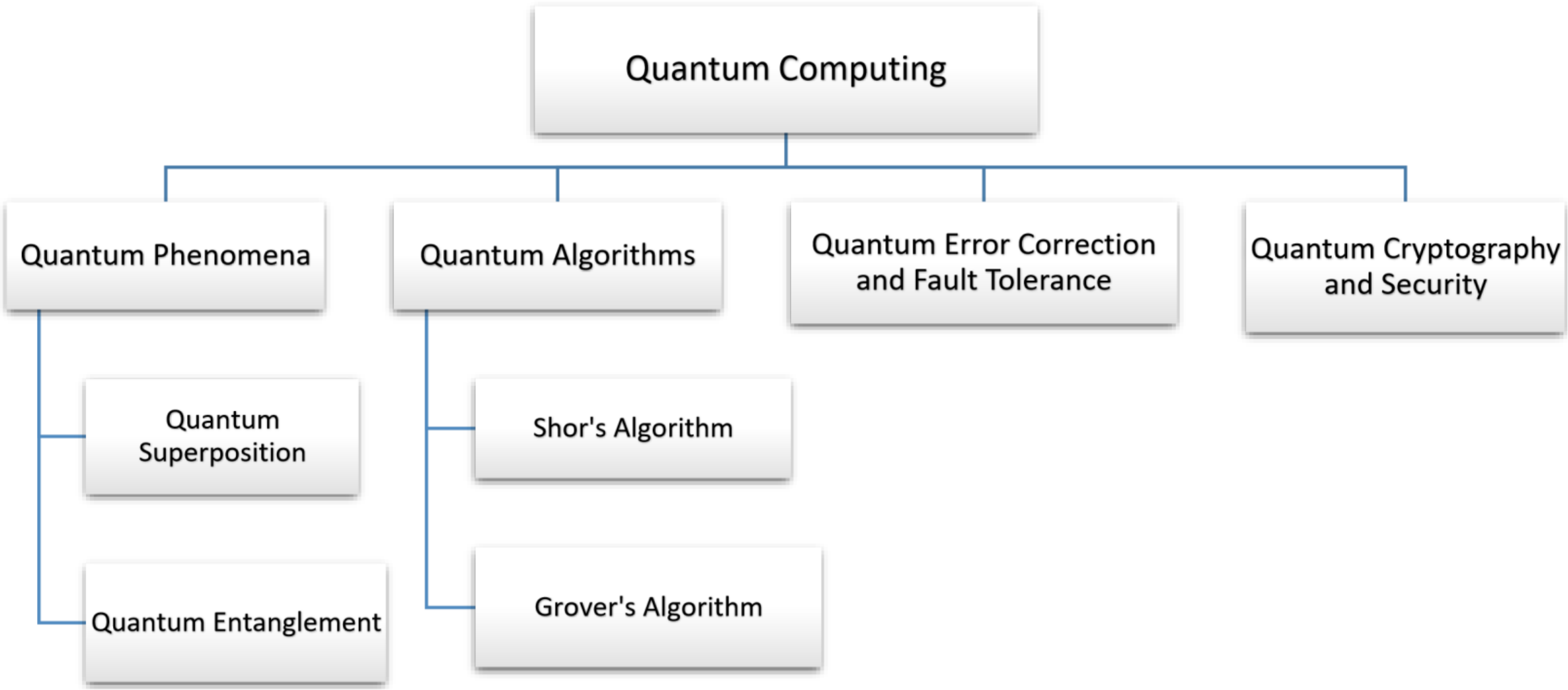

2018).

This paper will analyze four key components of quantum computing, as illustrated in Figure 2.

*Figure 2. Classification of quantum computing*

1. Quantum phenomena

They are physical effects that arise from the principles of quantum mechanics, such as superposition and entanglement. These effects occur at the atomic and subatomic levels, which are used by quantum computers to solve complex problems.

a. Quantum Superposition

Superposition is the property that allows qubits to exist in multiple states simultaneously. Mathematically, this corresponds to a principle of the Schrödinger equation, where any linear combination of its solutions is also a valid solution (Dey et al., 2022). This enables quantum computers to perform computations on all possible states of the qubits at once, exponentially increasing the number of values processed in a single operation. An *n*-qubit quantum computer can thus handle up to $2^n$ parallel operations. However, a qubit remains in superposition only in the absence of measurement. Upon observation, the quantum state collapses to one of the possible basis states (Mavroeidis et al., 2018).

b. Quantum Entanglement

Even more intriguing than superposition is the phenomenon referred to by Einstein as "spooky action at a distance," described in the EPR (Einstein–Podolsky–Rosen) paradox. This gave rise to the concept of quantum entanglement, offering a more profound and more complex interpretation of quantum mechanics (Dey et al., 2022). When two qubits become entangled, their states are no longer independent and must be described as a unified system with four possible combined states. Any change to the state of one qubit instantaneously affects the state of the other, regardless of the distance separating them. This phenomenon enables genuine parallel processing capabilities in quantum computing (Mavroeidis et al., 2018). In classical computing, increasing the number of bits results in a linear rise in processing capability. In contrast, quantum entanglement allows the addition of each qubit to enhance the computational power of a quantum system exponentially.

Thus, the quantum phenomena provide quantum computers an exponential parallel processing power which helps the quantum algorithms to break the current encryption systems causing threat to the cryptocurrencies using blockchain.

2. Quantum Algorithms

Quantum algorithms are computational procedures designed to run on quantum computers by leveraging quantum mechanical principles such as superposition and entanglement. There are many types of quantum algorithms depending on their application. Out of them, Shor's algorithm and Grover's algorithm demonstrate the potential threat to classical cryptography.

a. Shor's Algorithm

Shor's algorithm was published in 1994 by Peter Shor in his paper "Algorithms for Quantum Computation: Discrete Logarithms and Factoring" (Shor, 1994). The algorithm could factor large integers or solve discrete logarithmic problems faster with the help of a quantum computer (Mavroeidis et al., 2018). This led to a complete turnaround in quantum computing, significantly impacting various sectors such as cryptography, secure communications, optimization etc. Its foundation is based on period finding using Quantum Fourier Transformation (QFT). It can factor an integer N in time O($\log^3 N$) and space O(log N) (Kearney & PerezDelgado, 2021). It offers an exponential speedup in factoring integers compared to classical algorithms.

Shor's algorithm can also be applied to solve discrete logarithm problems. Vazirani (1994) elaborated on the procedure, demonstrating that starting from a random quantum superposition of two integers and applying a sequence of quantum Fourier transforms, one can construct a new superposition state that, with high probability, yields two integers satisfying a specific relation. This relation enables the determination of the period r, which corresponds to the unknown exponent in the discrete logarithm problem (DLP) (Mavroeidis et al., 2018).

b. Grover's Algorithm

Grover's algorithm aims to solve the problem of searching unstructured data by computing with high probability a unique (or very rare) solution *x* for which *f*(*x*) equals *v*, some desired value (Stewart et al., 2018).

A quantum computer using Grover's algorithm can find a specific entry in a database of N entries in $\sqrt{N}$ Searches as compared to N/2 searches required by a conventional computer (Mavroeidis et al., 2018; Grover, 1996). This helps a quantum computer to find the pre-image of a hash function faster than a classical computer and offers a quadratic speedup (Fernandez et al, 2021).

3. Quantum Error Correction and Fault Tolerance

While quantum computers hold significant promise as the foundation of faster and more advanced computational technologies, they are not without limitations. One of the most critical challenges lies in the susceptibility of qubits to errors, primarily caused by external influences such as environmental noise and thermal fluctuations. These factors significantly impair the stability and reliability of quantum computations.

To address these hardware-induced limitations and make a quantum computer scalable enough to execute quantum algorithms, a specialized field known as quantum error correction and fault tolerance has emerged, focusing on mitigating errors and ensuring reliable quantum computation. It is possible to reuse certain specific classical error-correcting schemes in quantum error correction (QEC). However, due to the no-cloning theorem, which states that an arbitrary quantum state cannot be copied exactly, it is difficult to translate classical codes into quantum ones (Roffe, 2019). Hence, QEC must use entanglement and other indirect measurements.

Thus, various different error correcting codes have been proposed such as Shor Code, Steane Code, CSS Code (Calderbank-Shor-Steane Codes), Surface Code etc (Shor, 1995; Steane, 1996; Calderbank & Shor, 1996; Fowler et al., 2012). Among these various types of error-correcting codes, the surface code is currently considered the most practical and robust due to its high threshold for error tolerance and compatibility with two-dimensional qubit architectures. Different quantum computing architectures employ various codes, but surface codes are the dominant choice in leading designs due to their scalability and fault tolerance (Webber et al., 2022).

The implementation of these codes plays a crucial role in extending decoherence time—the interval during which qubits can maintain their quantum state without degradation. To ensure effective error correction, a concept known as code cycle time—the time it takes to detect and correct an error—must be kept sufficiently short relative to decoherence time. High-fidelity quantum gates are also crucial for minimizing operational errors. These considerations are necessary to satisfy the threshold theorem, which asserts that as long as physical error rates remain below a certain threshold, arbitrarily long quantum computations are theoretically possible with sufficient error correction (Webber et al., 2022).

| Quantum Technology | Coherence Time | Gate Fidelity | Operation Speed | Scalability |
|---|---|---|---|---|
| **Superconducting** | 50–100 µs | 99.4% | 10–50 ns | Highly scalable |
| **Ion Trap** | >1000 s | 99.9% | 3–50 µs | Medium |
| **Photonics** | ~150 µs | 98% | ~1 ns | Highly scalable |
| **Neutral Atoms** | >1000 s | 95% | TBC | High |
| **Silicon-Based** | 1–10 s | 99% | 1–10 ns | Expected to scale well |
| **Topological Qubits** | Very high | Very high | Unknown | Medium - High |

*Table 3. Types of qubits used by quantum computers. Adapted from Assessment of Quantum Threat to Bitcoin and Derived Cryptocurrencies by S. Holmes & L. Chen, 2021, IACR Cryptology ePrint Archive, Report No. 2021/967. https://eprint.iacr.org/2021/967. Copyright 2021 by the authors.*

Currently, the quantum computers are in the NISQ era ie. Noisy Intermediate State Quantum (Stephen Holmes & Liqun Chen, 2021). Since qubits are susceptible to state changes and information loss from errors, a single logical qubit is encoded using multiple physical qubits. This added redundancy enables the system to identify and correct certain types of errors (Kappert, Karger, & Kureljusic, 2021). Thus, qubits used in various quantum computers are classified based on several factors, including the type of error-correcting codes employed, code cycle time, gate fidelity, and scalability, as shown in Table 3.

Thus, IBM developed a metric called Quantum Volume to measure the overall performance of a quantum computer. It takes into account several factors like the number of qubits, gate fidelity, connectivity, and error rates, not just qubit count. IBM created it to provide a realistic benchmark for comparing different quantum systems, focusing on how well they can run complex quantum circuits rather than just their size (Stephen Holmes & Liqun Chen, 2021).

4. Quantum Cryptography and Security

Quantum Cryptography aims to enhance communication security using the principles of quantum mechanics. Unlike classical cryptographic techniques that depend on computational hardness assumptions, quantum cryptography leverages the fundamental laws of physics to detect and prevent eavesdropping.

A prominent application in this domain is Quantum Key Distribution (QKD), which enables two parties to securely generate and share a secret key. Among the various QKD protocols, the BB84 protocol, introduced by Bennett and Brassard in 1984, remains the most widely implemented and studied due to its simplicity and proven security. Other forms of quantum cryptographic protocols include device-independent QKD, quantum digital signatures, and quantum secret sharing, all of which aim to future-proof secure communications in the presence of both classical and quantum threats.

QKD is not fully scalable today because ground-based key exchanges using optical fibers are limited to a few hundred kilometers due to the degradation of the quantum states containing the keys. Therefore, quantum blockchain networks leveraging quantum communication protocols will have to wait for a global QKD-based Internet, which is still a distant goal and cannot be relied upon for short-term quantum resistance (Allende et al., 2023).

### 3.1.2 QUANTUM THREATS TO CRYPTOCURRENCIES

The public key cryptography used by cryptocurrencies is asymmetric, which uses prime number factorization or the discrete logarithmic problem (Mavroeidis et al. 2018). This public key cryptography was earlier considered to be unbreakable; however, Shor's algorithm can easily factor large numbers or solve discrete logarithmic problems (Kappert, Karger, & Kureljusic, 2021).

On the other hand, the hash function used by cryptocurrencies in the consensus mechanism is a one-way function, ie, the input cannot be feasibly derived from the output. However, Grover's algorithm provides a quadratic speedup to calculate the inverse of a hash function (Kappert, Karger, & Kureljusic, 2021).

Thus, the Shor's algorithm and Grover's algorithm pose a threat to the cryptocurrencies which is classified as shown in figure 2.

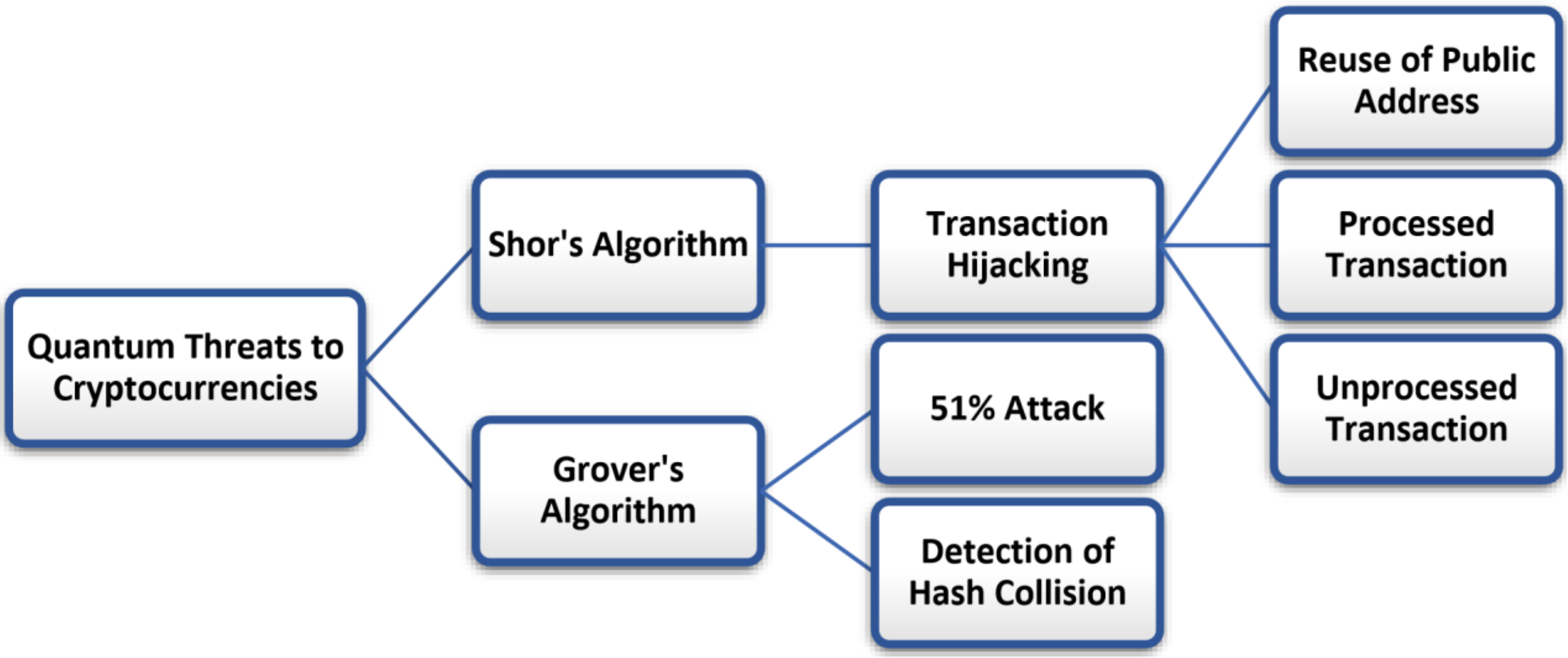

*Figure 2. Classification of Quantum Threats to Cryptocurrencies*

A. Threat due to Shor's Algorithm

Shor's Algorithm poses a threat to the public-key cryptography used by cryptocurrencies, specifically in the form of transaction hijacking (Stewart et al., 2018). In a transaction hijacking scenario, if an attacker employs Shor's algorithm, they could extract the victim's private key from their public key. This would allow the attacker to generate and broadcast a conflicting transaction using the same funds before the original transaction is confirmed in a block. By attaching a higher transaction fee, the attacker increases the chances that miners will prioritize and validate their fraudulent transaction over the legitimate one (Stewart et al., 2018; Kappert, Karger, & Kureljusic, 2021).

However, the extent of the threat posed by Shor's algorithm to cryptocurrencies varies depending on their underlying models, specifically, whether users reuse public addresses and whether a transaction is confirmed or still pending, as analysed by Aggarwal et al. (2018) and Stephen Holmes & Liqun Chen (2021).

1. Reuse of Public Address

Shor's algorithm poses a threat only if the public key is revealed. Once the public key is revealed in the presence of a quantum computer, the address is no longer safe. If a user reuses their public key, it is easy for the attacker to derive the private key and create fraudulent transactions. Thus, a user-based change is required to avoid reusing a public key more than once, thereby preventing a quantum attack.

2. Processed Transaction

If a transaction is conducted using a previously unused public key and has already been confirmed on the blockchain—validated by the majority of nodes—it is considered secure and finalized. In such a scenario, an attacker cannot use Shor's algorithm to derive the private key and alter the transaction. The only remaining possibility for an attack would be to outpace the network's hashing power and attempt a double-spending attack, potentially using Grover's algorithm to gain a computational advantage. Thus, processed transactions are considered to be reasonably secure.

3. Unprocessed Transaction

A transaction that has been broadcast to the network but is still pending in the mempool—i.e., not yet included in the blockchain—is at the highest risk. If an attacker manages to derive the private key from the public key before the transaction is confirmed, they can create a conflicting transaction using the same funds. By ensuring their transaction is added to the blockchain before the original one, the attacker can successfully redirect and steal the associated cryptocurrency. Thus, an unprocessed transaction is most vulnerable to Shor's algorithm while a processed transaction is fairly safe against it.

| Cryptocurrency | Public Key Reuse | Processed Transactions | Unprocessed Transactions |
|---|---|---|---|
| **Bitcoin** | High risk – addresses using P2PK expose public keys directly; reused addresses are vulnerable to private key extraction. | Low risk – once confirmed, secure unless attacker can perform 51% or Grover-based attack. | High risk – if the public key is visible before confirmation, the attacker can derive the private key and broadcast a conflicting transaction. |
| **Ethereum** | Very high risk – public keys are revealed as part of accountbased model even before spending, making all addresses potentially vulnerable. | Low to moderate risk – confirmed transactions are secure, but quantum attacks may affect smart contracts with exposed keys. | High risk – since keys are visible, attackers can act quickly before inclusion in a block. |
| **Litecoin** | Similar to Bitcoin – public key is exposed when funds are spent; reused keys are highly vulnerable. | Low risk – confirmed transactions are not easily reversible. | High risk – attacker can craft a double-spend using stolen keys if done before block inclusion. |
| **Zcash** | Moderate risk – shielded transactions (z-addresses) hide public keys; however, transparent addresses (taddresses) are like Bitcoin and vulnerable if reused. | Low risk – shielded transactions confirmed on-chain are secure; transparent processed ones follow Bitcoin's risk. | Moderate to high risk – shielded mempool is private, but transparent ones can be targeted similarly to Bitcoin. |
| **Monero** | Low risk – uses stealth addresses and ring signatures, preventing public key exposure and linking. | Very low risk – strong privacy prevents key visibility even post-confirmation. | Very low risk – unprocessed transactions do not reveal usable public key information. |

*Table 4. Vulnerabilities of major individual cryptocurrencies to Shor's algorithm*

However, it is essential to understand that the susceptibility of cryptocurrencies to quantum attacks is not uniform but varies based on the specific version of the public key address employed. Different address formats—such as legacy, hashed, script-based, and privacy-enhancing types—face varying levels of quantum vulnerability due to Shor's algorithm. While general risks can be outlined, the actual vulnerability in any given case depends on whether the address type exposes the public key directly, hashes it, or conceals it until the transaction is processed. Therefore, any assessment of quantum risk must account for the particular address version in use, as the level of security changes accordingly across and within cryptocurrencies as shown in table 5 (Kearney & Perez-Delgado, 2021).

| Public Key Version / Address Type | Used By | Explanation of version of public key | Reuse of Public Key Risk | Unprocessed Transaction Risk | Processed Transaction Risk |
|---|---|---|---|---|---|
| **Pay to Public Key Hash (P2PKH)** | Bitcoin, Litecoin | Hash of public key; public key revealed after spending | High | High | Medium |
| **Pay to Script Hash (P2SH)** | Bitcoin, Litecoin | Hash of redeem script; public key exposed when script executes | High | High | Medium |
| **Pay to Witness Public Key Hash (P2WPKH)** | Bitcoin | SegWit format; separates signature from transaction data | High | High | Medium |
| **M-address (Litecoin-specific prefix)** | Litecoin | Alternative prefix for SegWit; functionally similar to P2WPKH | High | High | Medium |
| **Externally Owned Account (EOA)** | Ethereum | The address is directly derived from the public key and is revealed with every transaction. | High | High | High |
| **Transparent address** | Zcash | Works like Bitcoin addresses; no privacy or encryption | Medium–High | High | Medium–High |
| **Shielded address** | Zcash | Uses zk-SNARKs to hide sender, receiver, and amount | Low | Low | Low |
| **Stealth address + Ring Signature** | Monero | One-time addresses with ring members hide the real sender; pk is never exposed | Low | Low | Low |

*Table 5. Vulnerability of different versions of public key address to Shor's algorithm*

Therefore, to break existing public key encryption systems, a quantum computer must be sufficiently advanced and fast enough to decrypt the data before the cryptocurrency transaction is confirmed and added to the blockchain. The number of physical qubits required to break an encryption by a quantum computer depends on several issues, including the efficiency of fault-tolerant error correcting codes, the physical error models and error rates of the physical quantum computer, optimizations in quantum factoring algorithms, and the efficiency of the synthesis of factoring algorithms into fault-tolerant gates (Mosca, 2015).

Based on the logical and physical resource estimates originally discussed by Häner et al. (2020) and Aggarwal et al. (2018), Webber et al. (2022) conducted a detailed analysis to determine the physical qubit requirements for a quantum computer capable of breaking Bitcoin's elliptic curve encryption. Assuming the use of surface code error correction, with a code cycle time of 1 µs, a reaction time of 10 µs, and a physical gate error rate of $10^{-3}$, the study estimated that approximately. $317 \times 10^6$ physical qubits would be required to break the encryption within one hour. For a 24-hour attack window, the requirement reduces to $13 \times 10^6$ physical qubits, while an attack completed in 10 minutes would demand around $1.9 \times 10^9$physical qubits.

## B. Threat due to Grover's Algorithm

### 1. 51% Attack

Grover's algorithm poses a significant threat to blockchain security by enabling a form of attack known as a 51% attack, where an attacker controls more than half of the network's computing power (Kappert, Karger, & Kureljusic, 2021). According to Nakamoto (2008), the Bitcoin protocol treats the longest chain as the valid one, primarily because not all nodes in the network hold identical local copies of the blockchain, often due to network delays or malicious interference. In classical computing, it is computationally infeasible for an attacker to outpace the entire network's computational power, usually after six blocks have been appended (Aggarwal et al. (2018)). However, Grover's algorithm reduces the complexity of breaking a k-bit hash function in just $2^{\frac{k}{2}}$ tries as compared to $2^k$ tries required by a classical computer, significantly accelerating the process (Rodenburg & Pappas, 2017). This quantum advantage enables an attacker to outperform the network's consensus mechanism by generating blocks at a faster rate, potentially reconstructing the blockchain and outpacing the legitimate chain maintained by honest nodes (Dey et al., 2022; Cui et al., 2020). Furthermore, the attacker can also prevent his own spending transactions to be recorded on the blockchain (Kappert, Karger, & Kureljusic, 2021).

While Grover's algorithm theoretically enables an attacker to perform a 51% attack by accelerating the process of finding valid hashes, this remains theoretical mainly under current technological limitations. In practice, for a quantum computer to successfully execute such an attack, it must achieve a particular hash rate that surpasses 50% of the total network's combined computational power. Thus, the performance of a quantum computer largely determines its ability to execute Grover's algorithm in the 51% attack (Kearney & PerezDelgado, 2021). According to Amy et al. (2016), a quantum computer can break the SHA-256 hash function commonly used in most cryptocurrencies with 14 million physical qubits or 2402 logical qubits.

### 2. Detection of Hash Collisions

In contrast to classical computers, quantum computers—through the application of Grover's algorithm—can significantly enhance the ability to invert cryptographic hash functions. This capability enables attackers to search for hash collisions. A hash collision occurs when two different inputs produce the same output hash value from a hash function. This allows the attacker to manipulate block content without disrupting the overall integrity of the blockchain (Fernández-Caramés & Fraga-Lamas, 2020). Usually, altering any data within a block changes its hash, rendering the block invalid and severing its link to the chain. However, if a malicious attacker can generate a collision—where a different set of data yields the same hash—they could modify the block's contents while preserving the chain's continuity, thereby compromising the system's security (Kappert, Karger, & Kureljusic, 2021; Kiktenko et al., 2018).

### 3.1.3 COUNTERMEASURES TO RESIST QUANTUM ATTACKS

Due to the quantum threats posed by Shor's and Grover's algorithms, which can potentially break widely used cryptographic schemes, many researchers and institutions have begun developing solutions to make blockchain systems resistant to quantum attacks. Among these, Post-Quantum Cryptography (PQC) stands out as the most actively researched and promising approach. Let us look at the possible countermeasures:

A. Post Quantum Cryptography

Since many classical cryptography schemes are considered insecure against quantum computers, researchers have introduced a field of study called the Post Quantum Cryptography (PQC) (Kappert, Karger, & Kureljusic, 2021). The Post post-quantum cryptographic algorithms are considered to be quantum-resistant and are mainly of the following types (Halak et al., 2024):

1. Hash-based

   Hash-based cryptography relies on cryptographic hash functions—non-reversible functions that produce fixed-length outputs from inputs of any size. Mainly used for digital signatures, these schemes are strong candidates for post-quantum security as they resist both classical and quantum attacks. They work by combining many One-Time Signatures (OTS) in a tree structure (like a Merkle tree), with each message signed using a unique OTS key and the tree root serving as the public key.

2. Lattice-based

   Lattice-based cryptography is a class of cryptographic schemes built on the hardness of lattice problems, such as the Shortest Vector Problem (SVP) and Learning with Errors (LWE). These problems are believed to be secure even against quantum computers, making lattice-based schemes strong candidates for post-quantum cryptography. These schemes support both encryption and digital signatures, and are known for their efficiency, simplicity, and resistance to quantum attacks.

3. Code-based

   Code-based cryptography refers to cryptographic systems built on the principles of error-correcting codes. In these systems, deliberate errors are introduced into messages to conceal their content, and only recipients with the appropriate private key can successfully correct these errors. The strength of this approach lies in the computational difficulty of decoding a corrupted codeword without knowing the underlying code's secret structure.

4. Multivariate-based

   Multivariate-based cryptography is a class of post-quantum cryptographic schemes that rely on the difficulty of solving systems of multivariate polynomial equations over finite fields, a problem considered hard even for quantum computers. These schemes are primarily used for digital signatures, and their main advantage is fast signing and verification. However, they often suffer from large key sizes and have faced security challenges over time, with some proposals being broken during cryptanalysis.

The PQC algorithms are considered to be NP-hard ie. quantum algorithms cannot easily break them (Gao et al., 2018). Additionally, post-quantum cryptographic (PQC) algorithms are designed to replace current digital signature and public key encryption schemes, protecting them from being broken by quantum computers.

In 2016, NIST launched the Post-Quantum Cryptography Standardization project (outlined in NISTIR 8105), with the goal of identifying and standardizing quantum-resistant cryptographic algorithms. The initiative focuses on replacing vulnerable encryption schemes with secure alternatives for core primitives such as digital signatures and key encapsulation mechanisms (KEMs) (Dey et al., 2022).

The process of selection was the following (Redkins, Kuzminykh, & Ghita, 2023):

- Round 1: 69 submissions – 20th December, 2017
- Reports of Round 1: 26 candidates selected – January, 2019
- Reports of Round 2: 15 candidates selected (7 finalists and 8 additional) – July, 2020
- Reports of Round 3: 4 algorithms are winners, and 4 candidates are selected for further process – July, 2022
- Reports of Round 4: 1 additional KEM finalised alongside the 4 winners – March 2025

After the 4th round of standardisation, the public-key encryption and key-establishment algorithm that will be standardized is CRYSTALS–KYBER, and the digital signatures that will be standardized are CRYSTALS–Dilithium, FALCON, and SPHINCS+. While four of the alternate key-establishment candidate algorithms will advance to a fourth round of evaluation: BIKE, Classic McEliece, HQC, and SIKE (NIST IR 8413).
Thus, based on 4th round, the finalized versions of selected algorithms have been formalized into federal standards. The modified version of CRYSTALS-Dilithium has been published as the Module-Lattice-Based Digital Signature Standard (FIPS 204), SPHINCS+ has been standardized as the Stateless Hash-Based Digital Signature Standard (FIPS 205), and CRYSTALS-Kyber has been issued as the Module-Lattice-Based KeyEncapsulation Mechanism Standard (FIPS 203). However, FALCON is still under development, which will be dubbed as FN-DSA (NIST IR 8545). On the other hand, from the alternate key-establishment candidate algorithms, HQC (Code-based) has been finalised as the additional key-establishment algorithm (NIST IR 8545) as discussed in the fourth-round status report of NIST, 2025.

| **Use Case** | **FIPS Standard Name** | **Original Algorithm Name** | **Type** |
|---|---|---|---|
| **Digital Signature - PRIMARY** | Module-Lattice-Based Digital Signature Standard (FIPS 204) | CRYSTALS-Dilithium | Lattice-based (Module-LWE) |
| **Digital Signature** | Stateless Hash-Based Digital Signature Standard (FIPS 205) | SPHINCS+ | Hash-based |
| **Key Encapsulation** | Module-Lattice-Based Key-Encapsulation Mechanism Standard (FIPS 203) | CRYSTALS-Kyber | Lattice-based (Module-LWE) |
| **Additional Key Encapsulation** | *To be announced (not yet assigned)* | HQC | Code-based |

*Table 6. NIST Standardised Algorithms*

Post-quantum cryptography (PQC) provides critical protection against quantum threats, but it presents several limitations to its adoption. Many PQC schemes require large key and signature sizes, which strain storage and bandwidth—especially in systems processing numerous transactions. Some schemes also suffer from slow key generation, limiting their scalability. High computational and energy demands make them difficult to deploy on existing blockchain infrastructure, particularly resource-constrained nodes. In addition, many PQC schemes remain unstandardized, creating risks of adopting algorithms that may later be broken or rejected. Large ciphertext overheads in some approaches further complicate practical integration. These challenges must be resolved for PQC to be effectively and widely adopted in blockchain systems (Fernández-Caramés and FragaLamas, 2020).

B. Post-quantum blockchain

As discussed by Fernández-Caramés and Fraga-Lamas (2020), various researchers and organizations are working to adapt cryptocurrency blockchains to resist the threats posed by quantum computers. These efforts involve integrating post-quantum cryptographic techniques into existing blockchain architectures or proposing entirely new quantum-resistant cryptocurrency frameworks. For instance, enhancements to major cryptocurrencies, such as Bitcoin and Ethereum, include replacing the vulnerable ECDSA signature scheme with quantum-secure alternatives, such as those based on hash functions like SHA-3 or BLAKE2.

Some approaches also explore quantum-safe key exchange protocols, such as Supersingular Isogeny Diffie–Hellman (SIDH), and signature schemes, including Rainbow and Niederreiter's code-based cryptosystem. Other cryptocurrency projects, like IOTA's Tangle, aim for quantum resistance through the use of one-time hash-based signatures and research into ternary computing hardware. Additionally, newer blockchains are incorporating post-quantum schemes, such as XMSS, designed to replace classical elliptic curve methods.

These advancements help protect cryptocurrencies from potential quantum attacks that could otherwise compromise private keys, forge signatures, and reverse transactions, thus preserving the integrity and trust essential to decentralized systems.

C. Quantum-secured Blockchain

Based on the properties of quantum mechanics, a quantum-secured blockchain can be created using quantum cryptography. It is one of the only long-term solutions to the encryption against the threat by quantum algorithms (Allende et al., 2023). It uses quantum cryptography for secure communication as explained above in section (3.1.2) (B)(4). Among various quantum cryptographic techniques, Quantum Key Distribution (QKD) is considered the most relevant and practical for enhancing encryption security. It provides two significant advantages: it generates completely random keys that are resistant to both classical and quantum attacks, ensuring forward secrecy, and it leverages the principle that any attempt to measure a quantum state disturbs it—making eavesdropping immediately detectable (Kappert, Karger, & Kureljusic, 2021).

Rajan and Visser (2019) describe a quantum blockchain that utilizes entanglement in time, where the functionality of time-stamped blocks and hash functions is replaced with a temporal GHZ state, enabling the linking of blocks through quantum temporal correlations rather than classical cryptographic methods. This can significantly be used in cryptocurrencies to enhance its security.

D. User-based and Soft Fork Changes

1. User-based changes (Stephen Holmes & Liqun Chen, 2021):-

i. Never re-use a public key address as it reduces the risk a quantum attack using Shor's algorithm
ii. Use multi-signatures as it can increase the number of qubits required by a quantum computer and therefore increase the cost of an attack
iii. Users can reduce attractiveness to a quantum attacker by holding a maximum value of cryptocurrency in each address that would make a quantum adversaries attack unprofitable.

2. Soft Fork Changes

   To address the dominance of ASIC miners, researchers have proposed alternative proof-of-work algorithms that emphasize memory intensity over raw computational speed. These include Momentum, which is based on finding collisions in hash functions (Larimer, 2014); Cuckoo Cycle, which focuses on detecting fixed-size subgraphs in random graphs (Tromp, 2015); and Equihash (Biryukov & Khovratovich, 2017), which leverages the generalized birthday problem. These algorithms are designed to be more democratic by reducing the efficiency gap between general-purpose hardware and specialized mining rigs. Additionally, their structure offers better resistance to quantum acceleration, making them promising options for future-proofing cryptocurrencies. Integrating such proof-of-work schemes into existing blockchain networks may require consensus rule changes; however, they can be introduced via soft forks, ensuring backward compatibility and avoiding network splits while gradually upgrading the system for improved fairness and quantum resilience (Aggarwal et al., 2018).

   Another important modification involves increasing the length of the public key, which raises the computational effort required by a quantum computer to carry out an attack using Grover's algorithm. As a result, the time needed to break the cryptographic scheme could exceed the block interval of the cryptocurrency, thereby mitigating the quantum threat. However, this change also impacts legitimate miners, as it increases the time needed to solve blocks—potentially leading to longer block intervals and reduced transaction throughput. (Kappert, Karger, & Kureljusic, 2021).

### 3.1.4 LIMITATIONS AND FUTURE WORK

Although Shor's and Grover's algorithms pose significant threats to current cryptocurrencies, they also come with inherent limitations that restrict their practical impact at present: -

3. The qubits in quantum computers are highly unstable and error-prone. They have to be kept in isolated places at a very low temperature, cooler than space (Kappert, Karger, & Kureljusic, 2021). Thus, current quantum computers are not yet fully scalable, as they are still in the NISQ era (Dey et al., 2022).
4. An attacker would look to hijack a transaction containing high cryptocurrency funds whose block interval time will be relatively less than that of other transactions. Due to that, the cost of the attack is very high, serving as a practical limitation for adversaries, especially in the current era where scalable quantum computers are not yet widely available. Stephen Holmes & Liqun Chen (2021) analyse the cost of an attack by an adversary in their paper "Assessment of Quantum Threat to Bitcoin and Derived Cryptocurrencies".
5. The speed of current quantum computers is same as compared to ASIC mining computers (Aggarwal et al, 2018). Due to this, the attacker cannot perform a 51% attack and outpace the entire network by executing the consensus mechanism faster, given its relatively slow speed.

However, if the gate speed of quantum computers reaches up to 100 GHz, it will be able to perform calculations 100 times faster than current technology, posing a plausible threat (Aggarwal et al., 2018).

Thus, current quantum computers face significant limitations such as instability, high cost, and limited speed. These challenges make real-world quantum attacks impractical for the time being. However, with ongoing advancements, the threat may become feasible in the future, making the shift to quantum-resistant cryptography increasingly important.

## 5. CONCLUSION

Quantum computing introduces a paradigm shift in computational capability, posing both profound threats and opportunities for blockchain-based cryptocurrencies. As analyzed, Shor's algorithm jeopardizes publickey cryptographic schemes by enabling the extraction of private keys, while Grover's algorithm weakens the resilience of hash functions used in consensus mechanisms, thereby facilitating attacks such as transaction hijacking and 51% consensus takeovers. Although current quantum computers lack the scalability, stability, and speed to pose immediate risks, rapid advancements in quantum hardware and error correction technologies make future threats a plausible concern.

To counteract these emerging challenges, two principal long-term solutions have emerged: Post-Quantum Cryptography (PQC) and Quantum Key Distribution (QKD). PQC provides quantum-resistant cryptographic schemes that can be integrated into existing blockchain frameworks through algorithmic upgrades or soft forks. Meanwhile, QKD leverages the principles of quantum mechanics to offer theoretically unbreakable encryption, although its practical scalability remains a limitation at present.

Furthermore, adopting user-based best practices—such as avoiding public key reuse, using multisignatures, and minimizing wallet exposure—can significantly reduce vulnerability in the near term. Combining these with protocol-level changes, such as memory-hard consensus algorithms and increased cryptographic key lengths, lays a robust foundation for quantum-resilient blockchains.

Ultimately, the convergence of quantum computing and blockchain necessitates proactive adaptation. Ensuring the security and longevity of decentralized financial systems will depend on timely standardization, implementation of PQC algorithms, and sustained research into scalable quantum-secure communication systems.

## 6. REFERENCES/CITATIONS